\documentclass{PoS}

\newcommand{\BlackHat}{{\sc BlackHat}}
\newcommand{\SHERPA}{{\sc SHERPA}}

\newcommand{\POWHEGBOX}{{\sc POWHEG BOX}}

\def\Zjj{$Z\,\!+\,2$}
\def\Zjjj{$Z\,\!+\,3$}
\def\Zjjjj{$Z\,\!+\,4$}

\def\Wjj{$W\,\!+\,2$}
\def\Wjjj{$W\,\!+\,3$}
\def\Wjjjj{$W\,\!+\,4$}
\def\Wjjjjj{$W\,\!+\,5$}
\def\Wpmjjjj{$W^{\pm}\,\!+\,4$}

\def\gjj{$\gamma\,\!+\,2$}
\def\gjjj{$\gamma\,\!+\,3$}

\newif\ifdraft
\drafttrue

\newif\ifpreprint
\preprinttrue

\title{
\ifpreprint
\hbox{\rm\small
UCLA/12/TEP/107$\null\hskip 3.7cm \null$
SLAC--PUB--15627$\hskip 4.5cm \null$ 
SB-F-413-12$\null\hskip 2.4cm \null$
\break}
\hbox{\rm\small 
IPPP/12/76 $\null\hskip 5.4cm \null$
LPN12-117 $\null\hskip 3.2cm \null$
 CERN--PH--TH/2012/279\break}
\hbox{$\null$\break}
\fi
High multiplicity processes at NLO with BlackHat and Sherpa}

\ShortTitle{High multiplicity processes at NLO with BlackHat and Sherpa}

\author{Zvi Bern, Kemal Ozeren\\
        Department of Physics and Astronomy, UCLA, 
		Los Angeles, CA 90095-1547, USA\\
        E-mail: \email{bern@physics.ucla.edu}, \email{ozeren@physics.ucla.edu}
		}
\author{Lance J. Dixon, Stefan H\"oche\\
        SLAC National Accelerator Laboratory, Stanford University,
        Stanford, CA 94309, USA\\
        E-mail: \email{lance@slac.stanford.edu}, \email{shoeche@slac.stanford.edu}
		}
\author{Fernando Febres Cordero\\
  		Departamento de F\'{\i}sica, Universidad
		Sim\'on Bol\'{\i}var, Caracas 1080A, Venezuela\\
        E-mail: \email{ ffebres@usb.ve}
		}
\author{Harald Ita$^*$\\
		Albert-Ludwigs-Universit\"at Freiburg, 
		Physikalisches Institut,
		D-79104 Freiburg, Germany\\
        E-mail: \email{harald.ita@physik.uni-freiburg.de}
		}
\author{David Kosower\\
        Institut de Physique Th\'eorique, CEA--Saclay,
        F--91191 Gif-sur-Yvette cedex, France\\
        E-mail: \email{david.kosower@cea.fr}
	}
\author{\speaker{Daniel Ma\^{\i}tre}
	\\
        Theory Division, Physics Department, CERN,
		CH--1211 Geneva 23, Switzerland\\
        Department of Physics, University of Durham,
          DH1 3LE, UK\\
        E-mail: \email{daniel.maitre@durham.ac.uk}
}

\abstract{ 
  In this contribution we review recent progress with
  fixed-order QCD predictions for the production of a vector boson in
  association with jets at hadron colliders, using the programs
  \BlackHat{} and \SHERPA{}.  We review general features of
  next-to-leading-order (NLO) predictions for the production of a
  massive vector boson in association with four jets.  We also discuss
  how precise descriptions of vector-boson production can be applied
  to the determination of backgrounds to new physics signals.  Here we
  focus on data-driven backgrounds to a missing-energy-plus-jets
  search performed by CMS.  Finally, we review recent progress in
  developing theoretical tools for high-multiplicity loop-computation
  within the \BlackHat{}-library.  In particular, we discuss methods
  for handling the color degrees of freedom in multi-jet predictions
  at NLO.}

\FullConference{
Loops and Legs in Quantum Field Theory - 11th DESY Workshop on Elementary
Particle Physics,\\ April 15-20, 2012\\ Wernigerode, Germany
}

\begin{document}
		\phantom{\speaker{H.~Ita and D.~Ma\^{\i}tre}}

\section{Introduction}

The production of a massive vector boson in association with jets is an
important process at hadron colliders.  The cross sections are large, the events
are relatively clean, and they form significant backgrounds to many interesting physics signals. 
Because these processes are so well understood, both experimentally and
theoretically, they are widely used to validate or test new tools and methods.

The last few years have seen continued progress in the perturbative
description of high-multi\-plicity processes.  The next-to-leading
order (NLO) QCD corrections for \Wjjjj{}-jet production at hadron
colliders were completed in 2011 \cite{W4,IOColor}, followed by the
calculation of the same process with a $W$ boson replaced by a $Z$
boson \cite{Z4}. At this conference we have shown preliminary results
for the \Wjjjjj{}-jet process. All these fixed-order QCD predictions
have been obtained using of \BlackHat{}~\cite{BH} and
\SHERPA{}~\cite{sherpa1,sherpa2,amegic1,amegic2}.  There has also been
a lot of progress in the computation of processes with a $W$ and $Z$
boson accompanied by jets at NLO accuracy, matched to a parton shower
(see e.g.~references in ref.~\cite{aMC@NLO_W2}). For such computations
the high-multiplicity virtual matrix elements in QCD are a key
ingredient, some of which have become available only recently.  The
production of \Wjj{} jets has been computed using aMC@NLO
\cite{aMC@NLO_W2}.  Virtual matrix elements provided by \BlackHat{}
have been used by different groups for such computations.  The Sherpa
implementation of the MC@NLO~\cite{MCatNLO} approach computes $W$
boson production in association to up to three jets \cite{SHERPA_W123}
using virtual matrix elements from refs.~\cite{W3PRL,W3}.  The
\POWHEGBOX{}~\cite{POWHEG1,POWHEG2} has been used recently to compute
the \Zjj{}-jet process at NLO, matched with a parton shower
\cite{POWHEG_Z2}, with the virtual matrix elements~\cite{ZqqQQ,Zqqgg}
also employed in ref.~\cite{Z3}.

NLO predictions improve the leading order (LO) results in
various ways.  NLO results show a reduced dependence on the unphysical
renormalization and factorization scales, as compared to LO results.  This
improvement becomes more significant as the number of jets and, thus, the order
in the strong coupling $\alpha_s$, increases.  
Further benefits include a better description of initial and
final state radiation. The shapes of many kinematical distributions are better
described at NLO. The precision offered by NLO calculations is needed for both
the signal and background processes.  `Data-driven' methods very often rely on theory input
for cross section ratios, in order to extrapolate backgrounds from a control region
into a signal region for the same process, or to extrapolate from one process to
a related process.   NLO computations can also improve the theoretical
precision for such ratios.

We address here several new developments in fixed-order NLO computations of
vector-boson production. First, we discuss the production of a massive vector
boson in association with four jets at NLO as a signal at the LHC. Next, we
discuss an application of NLO computations of vector-boson production to background
estimation in searches for supersymmetry.  Finally, we
report on some technical developments that have been important ingredients in
our latest multi-jet computations.

\section{\Wjjjj{}-jet and \Zjjjj{}-jet predictions}
\label{section:Z4}
We present results here for \Wjjjj{}-jet and \Zjjjj{}-jet
production, for which we have used \BlackHat{}~\cite{BH} for both the virtual
matrix elements and real-emission matrix elements.
\SHERPA{}~\cite{sherpa1,sherpa2,amegic1,amegic2} was used for the
remaining pieces (Born and subtraction), as well as for the
integration over the phase space. For both processes, the calculation
of the real-emission tree amplitudes is extremely challenging.  
In \BlackHat{}, they are computed using on-shell recursion relations
\cite{BCFW} as well as compact analytic formulae given in
refs.~\cite{DrummondAmplitudes,DHPS}.

\begin{figure}
\center{	\includegraphics[scale=0.5,clip]{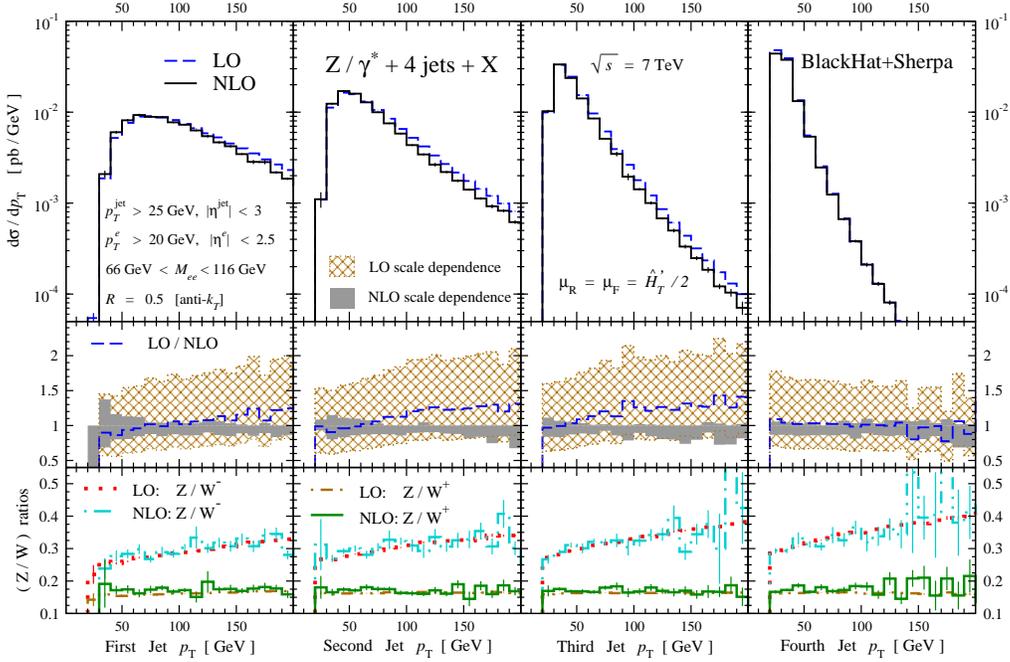}}
	\caption{Distribution in $p_T$ 
        for the first, second, third and fourth jets in
	\Zjjjj{}-jet events. }\label{fig:Z4} 
	\end{figure}

In figure \ref{fig:Z4} we display the transverse-momentum ($p_T$)
distribution of the first, second, third and fourth jets in
\Zjjjj{}-jet events, with the decay to a lepton pair included. We have
used $\hat{H}_T^{'}/2$ as a central choice for both the factorization
scale $\mu_F$ and renormalization scale $\mu_R$, where
$\hat{H}_T^{'}=E_T^Z+\sum_{i} p_T^i$ and
$E_T^Z=\sqrt{M_Z^2+\left(p_T^{e^+e^-}\right)^2}$. The sum runs over
all partons. A detailed list of the cuts, jet algorithm and parameters
can be found in ref.~\cite{Z4}.  To suppress the virtual-photon 
component, we apply a cut on the lepton invariant mass, requiring
$66\,\mbox{GeV}<M_{e^+e^-}<116\,\mbox{GeV}$.  Although the
interference between $Z$ boson and photon exchange is very small with
this cut, we include it for completeness.  For the virtual part we use
a leading-color approximation, which has been demonstrated in
refs.~\cite{W3,Z3,IOColor} to be good to about three percent, as
discussed further in section~\ref{sec:SLCimpact}.

In the top pane of figure \ref{fig:Z4}, the blue and black curves are
the central LO and NLO predictions, respectively.  In the middle pane
the ratio is taken with respect to the central NLO result.  Scale
variation bands are displayed in orange for the LO and in gray for the
NLO result.  They are obtained as the envelope from varying
$\mu=\mu_R=\mu_F$ by factors of $1/2$, $1/\sqrt{2}$, $1$, $\sqrt{2}$
and $2$ around the central scale value. The bottom pane displays the
ratio of the \Zjjjj{}-jet distributions with respect to the \Wpmjjjj{}-jet
ones, both at leading order and next-to-leading order.  In contrast to
the individual distributions, the ratios do not suffer from large
NLO corrections.  The shape of the $Z/W^-$ ratio is explained by the
dominance of the valence distributions, particularly $u(x)$, at large
$x$.  The $Z$ has a significant coupling to the $u$ quark, while the
$W^-$ couples only to the $d$ quark, among the valence quarks.  As the
jet $p_T$ increases, parton-distribution functions (PDF's) of higher
$x$ are probed, where $u(x)/d(x)$ rises.  The $Z/W^+$ ratio is flat
because both the $W$ and $Z$ couplings are dominated by the coupling
to the same $u(x)$ PDF.  These sorts of ratios are very useful in
data-driven methods.

The shape differs significantly between the central LO and NLO
predictions for the $p_T$ distributions of the first, second and third jets.
The scale variation is, as expected, much smaller at NLO than it is at LO.

\section{$Z/\gamma$ ratios}
The production of jets in association with a $Z$ boson that decays
into a neutrino pair is an irreducible background for missing energy
plus jets (METJ) searches.  Missing energy signals arise from the
production of new particles that escape the detector unobserved. A
typical example is the lightest supersymmetric particle (LSP) in
$R$-parity conserving models of supersymmetry.  As it is an
irreducible background, the impact of the
$Z(\rightarrow\nu{\nu})\,+\,$jets processes (METZJ) must be estimated
carefully.  Typically data-driven methods are used with only ratios
provided by theory. There are several strategies to estimate this
background (see
e.g.~refs.~\cite{CMSmemo,CMSMETJ1,StirlingZgam,CMSMETJ2}), using
related processes in which the production of the $Z$ boson
and its subsequent decay into neutrinos is replaced by a more
accessible signature:
\begin{itemize} 
\item $Z\,\,(\rightarrow l l)$+jets. The advantage of this process is
  that the production dynamics are very similar to the one in the
  neutrino process.  However, the statistics is smaller than
  $Z\rightarrow\nu\nu$ by a factor of six for each lepton flavor, and
  is further reduced by the experimental acceptance and identification
  efficiency for each charged lepton.
\item $W\,\,(\rightarrow l\nu)$+jets. This process has a cross section
  higher by a factor of six, but suffers from contamination from
  $t\bar{t}$, and potentially, from the new physics one aims to
  measure.
\item $\gamma$+jets. Replacing the $Z$ decay to neutrinos by a photon
  yields a cross section higher by a factor of four to five, but the
  production dynamics are different and a reliable theory prediction
  is needed to obtain a solid conversion factor.
\end{itemize}   

Here we present NLO calculations of the ratios needed for the photon-based estimation
strategy. The CMS collaboration has studied and used~\cite{CMSmemo,CMSMETJ1}
$W$-boson and photon production in association with jets to estimate METZJ
backgrounds~\cite{CMSMETJ1,CMSMETJ2}. Photon production in association with
jets has also been studied in ref.~\cite{StirlingZgam} and used by the ATLAS
collaboration~\cite{ATLASMETJ} in their data-driven estimates of the METZJ
background.

Partonic calculations involving photons develop infrared singularities
when the photon is emitted collinear to a quark. These divergences can
be avoided by imposing a standard cone isolation for the photon. However, this
strategy requires the use of photon fragmentation functions which are extracted
from data and are not known very precisely. Another approach, proposed by
Frixione \cite{FrixioneCone}, removes the need for the fragmentation functions
but is difficult to implement in an experimental measurement. For the
estimation of the ratio of \gjj{} jets to \Zjj{} jets we use the Frixione
isolation.  We estimate the difference between this isolation and the CMS cone
isolation using a code by Gordon and Vogelsang~\cite{Vogelsang} and
JetPhox~\cite{JetPhoX,JetPhoX2}. We find that the
difference between the standard cone isolation and Frixione's isolation is
small, and decreases as the transverse energy of the photon increases,
which is the case of interest for this study. For more details we refer to
ref.~\cite{gammaZ}.

\begin{figure}
\center{	\includegraphics[scale=0.4,clip]{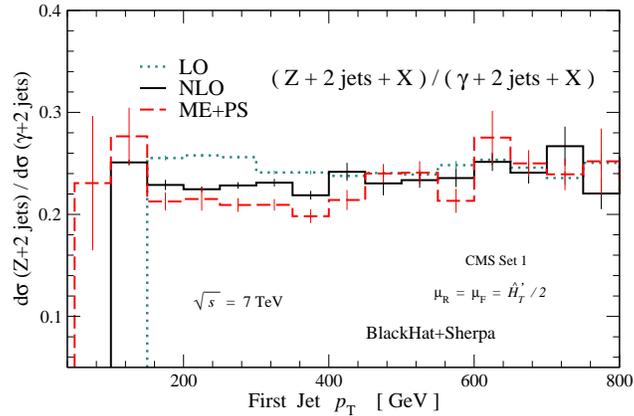}}
\caption{Ratio of the $p_T$ distribution of the first jet in \Zjj-jet and \gjj{}-jet production. The dotted blue line represents the LO ratio, the solid black line the NLO result, while the ME+PS result is represented by the dashed red line.  }\label{fig:zg} 
\end{figure}

The estimation of uncertainties on our result for the ratios (as for most NLO
predictions for ratios) is difficult. The commonly used method of varying the
renormalization and factorization scales yields a very small estimate, because most
of the scale-dependence is correlated between the numerator and
denominator. A similar effect occurs for the PDF-error propagation. To
estimate the error, we computed the ratio using the matrix-element-matched-to-parton-shower
(ME+PS) method of \SHERPA{}. We took the difference between the NLO and ME+PS ratios
as an estimate of the uncertainty of our result. An example of such a ratio, for cuts
relevant to a CMS search using the 2010 LHC data~\cite{CMSMETJ1}, is shown in Figure
\ref{fig:zg} as a function of the $p_T$ of the first jet.
The NLO and ME+PS predictions for the $Z$ to $\gamma$ ratio track each other
well across the whole range of jet $p_T$.  Other observables exhibit
similar behavior~\cite{gammaZ}, leading to the conclusion that the
photon-plus-jets process gives a good handle on the $Z\rightarrow \nu{\nu}$
background.

We have recently extended this computation \cite{gammaZ3} to the ratio of \Zjjj{}-jet
to \gjjj{}-jet production.  This work was also instigated by the CMS collaboration~\cite{CMSmemo}
to help them estimate the uncertainty in the $Z\rightarrow \nu{\nu}$ background to new
physics searches, in particular for the more stringent cuts used in analyzing the 2011 LHC
data~\cite{CMSMETJ2}.  We have computed the ratio of the two processes for
different sets of cuts on two kinematic variables $H_T^{\rm{jets}}$ and
$MET^{\rm{jets}}$.  $H_T^{\rm{jets}}$ is defined as the sum of the transverse
energy of jets with $p_T>50\;\rm{GeV}$ and $|\eta|<2.5$,
while $MET^{\rm{jets}}$ is defined as the modulus of the vectorial
sum of the transverse momenta of the jets with $p_T>30\;\rm{GeV}$ and
$|\eta|<5$. 

\begin{figure} \center{	\includegraphics[scale=0.4,clip]{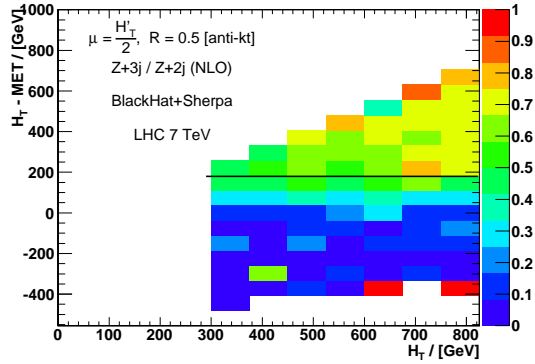}}
	\caption{Ratio of the \Zjjj{} to \Zjj{} jets ratio as a function of
	$H_T^{\rm{jets}}$ and $H_T^{\rm{jets}}-MET^{\rm{jets}}$.  }\label{fig:Z3Z2} 
\end{figure}

The scale $Q$ of the cuts on  $H_T^{\rm{jets}}$ and $MET^{\rm{jets}}$ in the
analysis is large enough that one may worry about large logarithms of the form
$\log(Q/p_T^{\rm{min}})$. One way of assessing whether these logarithms are
large is to look at the ratio of \Zjjj{} jets over \Zjj{} jets.  A value too
close to one could result from large logarithms spoiling the perturbative
expansion. Figure \ref{fig:Z3Z2} shows this ratio as a function of
$H_T^{\rm{jets}}$ and $H_T^{\rm{jets}}-MET^{\rm{jets}}$. In regions where this
ratio is close to one, we cannot be sure about the validity of the perturbative
expansion.
Such large effects may nonetheless largely
cancel out in $Z$+jets/$\gamma$+jets ratios,
leaving those predictions reliable in spite of them.
It is worth noting that the validity of the NLO prediction in these
regions, where the perturbative expansion might be unreliable, can be
tested experimentally by measuring such ratios in the
$\gamma\,+\,$jets samples.

We find that the theoretical uncertainty on the conversion between photons and $Z$ bosons
is less than 10\% for events with either two or three associated jets,
allowing the photon channel to provide an excellent determination of the Standard Model
missing-$E_T$\,+\,jets background.  

\section{Color Automation}

Scattering processes at hadron colliders are dominated by the
strong interactions. The previously-discussed processes of vector-boson
production in association with jets are typical examples. For events with
kinematics in the perturbative regime, cross sections and distributions can
be predicted from first principles, allowing detailed comparisons between
theory and experiment.  To this end we focus on the precise
perturbative description of the interactions of colored partons within
quantum-field theory.

The Lagrangian
interaction terms of colored fields make perturbative computations very
challenging.  A successful computational strategy disentangles the various
dynamical degrees of freedom, such as kinematics, spin
and color quantum numbers. Targeted algorithms can then be devised to deal
efficiently with the components of the computation.  This approach is
realized in many matrix-element generators and has been particularly useful
for recent multi-jet computations at NLO.
Alternatively, one can treat color simultaneously with the kinematic
variables~\cite{BGcolordressed}. This method, implemented as a Berends-Giele
recursion~\cite{BGrelations}, is similar to methods used in other automated LO
computations~\cite{alpha,helactree,othertree,COMIX}.

In \BlackHat{} we separate color and kinematic degrees of freedom early on. We
then compute color-ordered tree- and loop-level objects, so-called primitive
amplitudes. These amplitudes are assembled into the full matrix elements only
at the end of the computation. The required color weights and interference
matrix are precomputed.  We use modern on-shell and unitarity techniques in
order to compute the color-ordered building blocks: tree amplitudes and,
subsequently, loop amplitudes.

Focusing on these particular color-ordered objects has several benefits.
First of all, these components have a simpler analytic structure than the
full amplitude; for example, they depend on a reduced set of kinematic
invariants.  This property implies that a smaller number of unitarity cuts
needs to be computed in an on-shell approach.
The simpler analytic structure also leads to improved numerical stability
of the amplitudes. 

The second main benefit of the color-ordered approach is that it can be
exploited to yield significant efficiency gains in the numerical phase-space
integration~\cite{W3}.  If we consider the number of quark flavors $n_f$
to be of the same order as the number of colors $N_c$, and take the limit
as both become large, then the color-summed (virtual) cross section can be 
expanded in powers of $1/N_c^2$.   The $1/N_c^2$-suppressed terms are
numerically quite small.  Thus, for a fixed integration error, fewer
evaluations are needed for these parts of the cross section.  That is,
for most phase-space points only a small subset of all color-ordered
amplitudes must be computed, namely those that contribute to the leading-color
term in the cross section.

In this talk we discuss a method to automate the color-ordered approach in
loop computations. The key question is: how we can express generic matrix
elements as linear combinations of color-ordered objects.  We also discuss
the quantitative impact of subleading-color terms on differential cross
sections.  In particular, we consider the distribution of the fourth
jet $p_T$ in the state-of-the-art NLO results for \Wjjjj{}-jet
production~\cite{W3,W4,IOColor}.

\subsection{Partial amplitudes from primitive amplitudes.} 

The standard ``trace-based'' color decomposition of a one-loop QCD amplitude 
is into a set of color structures involving traces or open strings of
$N_c\times N_c$ $SU(N_c)$ generator matrices $T^{a_k}$, one
for each external gluon.  The open strings terminate on fundamental
indices corresponding to the external quarks and antiquarks in the process.
The coefficients of these color structures are called partial amplitudes.
The partial amplitudes are in turn built from color-ordered primitive
amplitudes, but the precise relations can be laborious to determine in the
general case.  The aim of this section is to describe an
algorithm~\cite{IOColor} for determining these relations.

For many specific processes, the decomposition of the one-loop matrix
elements and the color-summed virtual cross sections in terms of primitive 
amplitudes is well known.  The explicit color
decompositions of all-$n$ matrix elements of purely gluonic processes
can be found in refs.~\cite{BKColor,Neq4Oneloop} and the ones with a
single of quark line in ref.~\cite{TwoQuarkThreeGluon}.  Furthermore, the
decomposition of the four-quark
amplitude was given in refs.~\cite{KST4,ZqqQQ,Zqqgg} and that of the four-quark
process with an additional gluon in~\cite{KSTqqqqg,W3jcolordec}.
The decomposition of QCD scattering amplitudes with six and seven partons,
including either four or six quarks, have been given only
recently, using the algorithm reviewed here~\cite{IOColor}.

The algorithm is based on analyzing Feynman diagrams and
their inherent color information. We consider a specific subprocess in QCD. We
generate the full set of Feynman diagrams of the colored loop amplitude.
Next, we associate a linear combination of Feynman diagrams to partial
amplitudes.  To this end, we dress Feynman diagrams with color matrices and
sum over repeated (internal) color indices.  Partial amplitudes are defined
as coefficients of particular products of color matrices.  Feynman diagrams
are associated with a given partial amplitude if they contribute to its
defining color structure.  Simple factors (powers of $N_c$, signs and
integers) are generated by the color-index summation, and these
factors enter the relation between each Feynman diagram and each
partial amplitude to which it contributes.

In a second step, we associate a linear combination of Feynman diagrams with
primitive amplitudes. To do this, we again dress Feynman diagrams with
color matrices; however, this time we associate adjoint representation
color charges to quarks as well as gluons.  Again we sum over the internal
color indices.  We define the primitive amplitudes as the leading-color
single-trace partial amplitudes~\cite{BKColor} for a given cyclic ordering
of the external gluons and (adjoint) quarks.  A primitive amplitude is 
thus associated with the Feynman diagrams that contribute to a particular
single-trace color structure. For further details, and refinements when
dealing with quark lines, we refer the reader to the original
literature~\cite{BKColor,Neq4Oneloop,TwoQuarkThreeGluon,IOColor}.

Once we have the explicit expressions of both the partial amplitudes and the
primitive amplitudes in terms of linear combinations of Feynman diagrams,
we can manipulate these two sets of equations.  We express the partial
amplitudes in terms of the primitive amplitudes by solving the linear set
of equations to eliminate the explicit dependence on the individual
Feynman diagrams.  In this step, we observe a redundancy of the linear
equations we have to solve.  This redundancy implies that certain linear
combinations of primitive amplitudes add up to zero. That is, we find
non-trivial relations between primitive amplitudes. Thus, we may find a set of
equivalent color decompositions of a given partial amplitude.

We point out some key features of the implementation of the above algorithm.
Because it is based on Feynman diagrams, the algorithm is limited in
parton multiplicity due to the rapid growth in the number of diagrams.
(However, the algorithm only has to be performed once and for all for a 
given process, not each time an amplitude is evaluated.)
We find that there is no serious obstruction to carrying out the
decomposition
for processes with up to eight
external partons. Processes with zero or two external quarks are computationally
the most challenging; however, explicit formulas are known for these cases.
In addition, it turns out that many diagrams (e.g.~diagrams involving
four-gluon interactions) may be dropped from the start, significantly
reducing the computational load.  For an efficient implementation, we find
it convenient to use QGRAF~\cite{QGRAF} for diagram generation and the
computer algebra package FORM~\cite{FORM} for summing over internal
color indices.  

We conclude this section with a brief discussion of the relations between
primitive amplitudes, which appear as a byproduct of the above
algorithm. At low multiplicity, one can identify the anti-symmetry of
the color-ordered gluon-quark-quark vertex as the origin of the
relations between primitive amplitudes. Explicit examples of this may
be found in ref.~\cite{IOColor}. The relations obtained between
primitive amplitudes can be used to optimize caching, and thus,
how long the subleading-color contributions take to evaluate numerically.
Finally, it seems likely that understanding the relations between primitive
amplitudes will help to establish all-$n$ formulae for the color
decompositions.  We have not explored this direction further.

\subsection{Quantitative impact of subleading-color terms.} 
\label{sec:SLCimpact}

We now discuss the quantitative impact of subleading-color terms on NLO
predictions. We focus on the distribution in the fourth jet $p_T$
in \Wjjjj{}-jet production at the LHC, since it is a key observable.
Predictions of jet $p_T$-distributions have already been given in 
ref.~\cite{W4}. In that work a leading-color approximation was used for 
the virtual parts; the remaining real, Born and subtraction terms were
computed to all orders in the color expansion.
Here we show results including the remaining subleading-color
corrections~\cite{IOColor}.  The systematics we observe match our
earlier results on subleading-color contributions to \Wjjj{}-jet
production \cite{W3}.

We use the same basic setup of matrix-element generators, Monte Carlo
integration, scale choices and cuts as discussed in section~\ref{section:Z4}.
There is no unique definition of the leading-color terms; definitions may
differ by reassigning subleading-color terms.  A detailed discussion of the
leading-color approximation we use here is provided in ref.~\cite{IOColor}. 
\begin{figure} 
\includegraphics[scale=0.6]{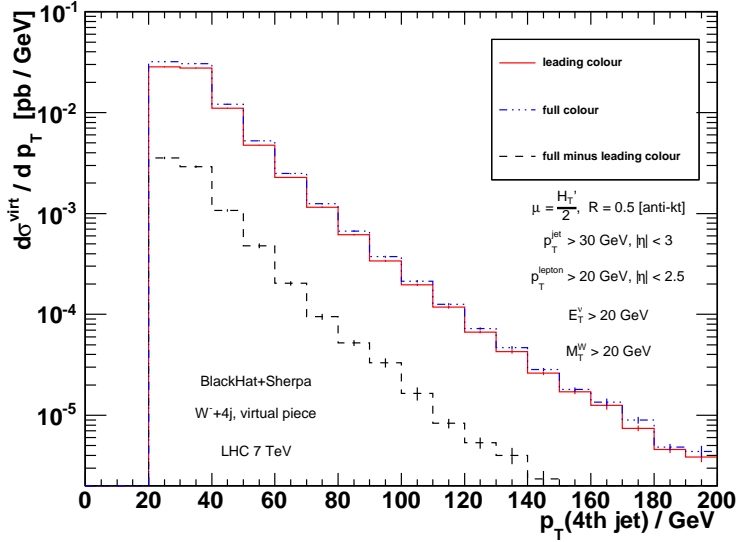}
\caption{ A comparison of the full and leading-color virtual contributions to
the $p_T$ distribution of the fourth jet in $W^-$+4-jet production at the
$7$~TeV LHC.}
\label{fig:Wm4j_virt_pt} \end{figure}

In figure~\ref{fig:Wm4j_virt_pt} we compare the full-color versus leading-color
virtual contributions to the $p_T$ distribution of the fourth jet in
$W^-$+4-jet production.  Also displayed are the subleading-color contributions
by themselves, labeled as `full-minus-leading-color'.  In order to obtain
the full parton level differential cross section one must add the real and Born
contributions in the usual way. The size of these contributions have been given
 recently~\cite{W4,W3}; the leading-color virtual part accounts for
about $20\%$ of the cross section.

Figure~\ref{fig:Wm4j_virt_pt} confirms that the subleading-color
contribution is suppressed uniformly by almost a factor of 10 
over the displayed range of $p_T$.  The suppression appears consistent with
the expected factor of $1/N_c^2$ with $N_c=3$.  With the leading-color
virtual part accounting for about $20\%$ of the total cross section,
the subleading-color virtual part amounts to less than a 3\% correction
to the total.

A possible exception to the uniform suppression would be near a zero of the
leading-color virtual cross section. Such zeros are not excluded on
general grounds, but they would have to survive the sum over a large number
of different helicity configurations.  A priori, there is no reason to
assume that the vanishing of the leading-color contribution forces also
the vanishing of the subleading-color contribution.
Close to a putative zero of the leading-color contribution we would expect a
relative enhancement of the subleading-color piece in the virtual matrix
elements.  Even if a zero were to appear in a special kinematic configuration,
the subleading-color virtual terms are expected to keep their relative size
with respect to the more inclusive total cross section and typical 
differential cross sections.  Certainly figure \ref{fig:Wm4j_virt_pt} does
not show any evidence for suppression of the leading-color cross section.

Although the leading-color results could point to kinematic
configurations where subleading-color contributions might dominate the
virtual matrix elements, only their explicit knowledge allows to
determine their impact on the full cross section. Similarly, the size
and uniformity of the subleading-color terms justifies the use of
leading-color approximations for many multi-jet observables.
Nevertheless, explicit computations are of great importance for a
definitive understanding of such multi-jet observables.

\section*{Acknowledgements}
DAK's work is supported by the European Research Council under Advanced
Investigator Grant ERC--AdG--228301. DM's work was supported by the Research
Executive Agency (REA) of the European Union under the Grant Agreement number
PITN-GA-2010-264564 (LHCPhenoNet). This research was supported by the US Department of Energy under contract DE--AC02--76SF00515 and DE--FG03--91ER40662.

\providecommand{\href}[2]{#2}
\addcontentsline{toc}{section}{References} \bibliographystyle{JHEP}
\bibliography{ref}

\end{document}